\documentclass[english]{article}
\usepackage[T1]{fontenc}
\usepackage[latin9]{inputenc}
\usepackage{color}
\usepackage{float}
\usepackage{graphicx}

\usepackage{babel}

\begin{document}

\title{The Impact of Past Epidemics on Future Disease Dynamics}

\author{Shweta Bansal%
\thanks{Center for Infectious Disease Dynamics, The Pennsylvania State University,
208 Mueller Lab, University Park PA 16802%
} %
\thanks{Fogarty International Center, National Institutes of Health, Bethesda,
MD 20892, USA %
} %
\thanks{Corresponding Author: shweta@sbansal.com%
}, Lauren Ancel Meyers%
\thanks{Section of Integrative Biology and Institute for Cellular and Molecular
Biology, University of Texas at Austin, 1 University Station, C0930,
Austin, TX 78712, USA %
} %
\thanks{Santa Fe Institute, 1399 Hyde Park Road, Santa Fe, NM 87501, USA%
}}
\maketitle
\begin{abstract}
Many pathogens spread primarily via direct contact between infected
and susceptible hosts. Thus, the patterns of contacts or \textit{contact
network} of a population fundamentally shapes the course of epidemics.
While there is a robust and growing theory for the dynamics of single
epidemics in networks, we know little about the impacts of network
structure on long term epidemic or endemic transmission. For seasonal
diseases like influenza, pathogens repeatedly return to populations
with complex and changing patterns of susceptibility and immunity
acquired through prior infection. Here, we develop two mathematical
approaches for modeling consecutive seasonal outbreaks of a partially-immunizing
infection in a population with contact heterogeneity. Using methods
from percolation theory we consider both \textit{leaky immunity},
where all previously infected individuals gain partial immunity, and
\textit{perfect immunity}, where a fraction of previously infected
individuals are fully immune. By restructuring the epidemiologically
active portion of their host population, such diseases limit the potential
of future outbreaks. We speculate that these dynamics can result in
evolutionary pressure to increase infectiousness. 
\end{abstract}

\section{Introduction}

Immunity acquired via infection gives an individual protection from
subsequent infection by the same or similar pathogen for some period
of time. For diseases such as measles, varicella (chickenpox), mumps
and rubella, complete immunity lasts a lifetime; therefore an individual
who has been infected by one of these pathogens, once recovered, cannot
be reinfected, nor transmit the infection again. For other diseases,
immunity wanes with time, leaving previously infected individuals
only partially protected against reinfection (called \textit{partial
immunity}). This degradation of immunity may be caused by antigenic
variation in the circulating pathogen or loss of antibodies over time.
The transition from complete to partial immunity can happen over different
timescales: over a few weeks as with norovirus and rotavirus \cite{rotavirus},
over months or a few years as with influenza \cite{flu_book}, or
over many years as with pertussis \cite{pertussis}. Here, we present
new methods for modeling the epidemiological consequences of partial
immunity. 

\textcolor{black}{Although partial immunity is not well-understood,
there is evidence that partial immunity functions in one of two ways:
leaky or perfect. For a degree of partial immunity $q$, leaky partial
immunity implies that each immunized individual reduces their chances
of getting reinfected and infecting others by a proportion $q$, whereas
perfect partial immunity implies that a fraction $q$ of immunized
individuals enjoy full protection from reinfection and the remaining
$\left(1-q\right)$ proportion are completely susceptible. Leaky partial
immunity is expected to be the more common of the two, and more consistent
with our understanding of the immune system \cite{immunology}. Perfect
partial immunity is less common, but can occur if some individuals
are unable to mount a lasting immune response to an otherwise fully
immunizing disease. It has been observed, for example, in vaccine
and animal studies for varicella, meningococcal infection \cite{varicella},
and Hepatitis C \cite{bukh,hepc_farci}.}

Partial immunity may impact the host in multiple ways, and have far-reaching
implications for the transmission of a disease through a population.
Specifically, it can decrease one or both of two fundamental epidemiological
quantities: \emph{infectivity}, the probability that an infected individual
will infect a susceptible individual with whom he or she has contact;
and \emph{susceptibility}, the probability that a susceptible individual
will be infected if exposed to disease via contact with an infected
individual. In mathematical models, the probability of transmission
(\textit{transmissibility}) during a contact between an infected and
susceptible individual is often represented as a product of the infectivity
of the infected node and the susceptibility of the susceptible node.
Partial immunity can limit transmissibility either by lowering the
probability of reinfection or reducing the degree to which an infected
individual sheds the pathogen. Both, for example, occur in the case
of influenza \cite{flu_immunity,flu_immunity_2}. 

Mathematical modeling of infectious disease dynamics has been dominated
by the Susceptible-Infected-Recovered (SIR) compartmental model \cite{kermack}
which considers infectious disease transmission in a closed population
of individuals who enjoy complete immunity following infection. The
SIR model has been extended to Susceptible-Infected-Recovered-Susceptible
(SIRS) dynamics to model the full loss of complete immunity after
a temporary period of protection \cite{hoppen,waltman}, and has been
applied successfully in several situations (e.g. \cite{grassly}).
Models of partially immunizing pathogens are less common, and have
primarily been developed for particular pathogens, such as influenza
\cite{recker_flu_immunity,levin_dushoff_plotkin,nuno_flu_models}.
They consider the impacts of antigenic variation and the resulting
complex patterns of cross-immunity on epidemic dynamics, but are limited
by the assumptions of homogeneous-mixing.

Contact network epidemiology is a tractable and powerful mathematical
approach that goes beyond homogeneous-mixing and explicitly captures
the diverse patterns of interactions that underlie disease transmission
\cite{barbour,watts,pastor_complex,meyers_sars,shirley,bansal_interface}.
In this framework, the host population is represented by a network
of individuals (each represented by a node) and the disease-causing
contacts (represented by edges) between them (Figure \ref{fig:network}(a)).
The number of contacts (edges) of a node is called its \textit{degree},
and the distribution of degrees throughout the network fundamentally
influences where and when a disease will spread \cite{meyers_sars,mejn,bansal_interface}.
The traditional SIR model has been mapped to a bond percolation process
on a contact network, in which individuals independently progress
through S, I, and R stages if and when disease reaches their location
in the network \cite{mejn}. The bond percolation threshold corresponds
to the epidemic threshold, above which an epidemic outbreak is possible
(i.e. one that infects a non-zero fraction of the population, in the
limit of large populations); and the size of the percolating cluster
(or giant component) above this transition corresponds to the size
of the epidemic. The standard bond percolation model for disease spread
through a network, however, assumes a completely naive population
without immunity from prior epidemics \cite{mejn}. 

In this paper, we extend the bond percolation framework to consider
the impact of infection-acquired immunity on epidemiological dynamics.
We model both perfect (Section 2.1) and leaky (Section 2.2) partial
immunity, and show that the two models are identical in the cases
of no immunity or complete immunity, but make very different predictions
for partial immunity. The evolution of infectiousness, virulence and
a pathogen's antigenic characteristics are in part driven by the epidemiological
environment. Although significant attention has been paid to the interaction
between contact network structure and pathogen evolution and competition
\cite{boots_sasaki,read_net_evolution,van_baalen,buckee_straindiversity,nunes_pathdiversity},
we do not yet understand the inter-seasonal interactions via modification
to the immunological structure of the host contact network. Feedback
from an evolving organism to its own ecological and evolutionary environment
is generally known as niche construction \cite{niche_feldman,boni_niche}.
Here, we use our models to explore a particular instance of niche
construction: the impacts of prior epidemics on the future dynamics
of the pathogen.

\section{Methods: Incorporating Infection-Acquired Immunity into a Network
Model}

\begin{figure}[h]
\begin{centering}
\includegraphics[width=11cm]{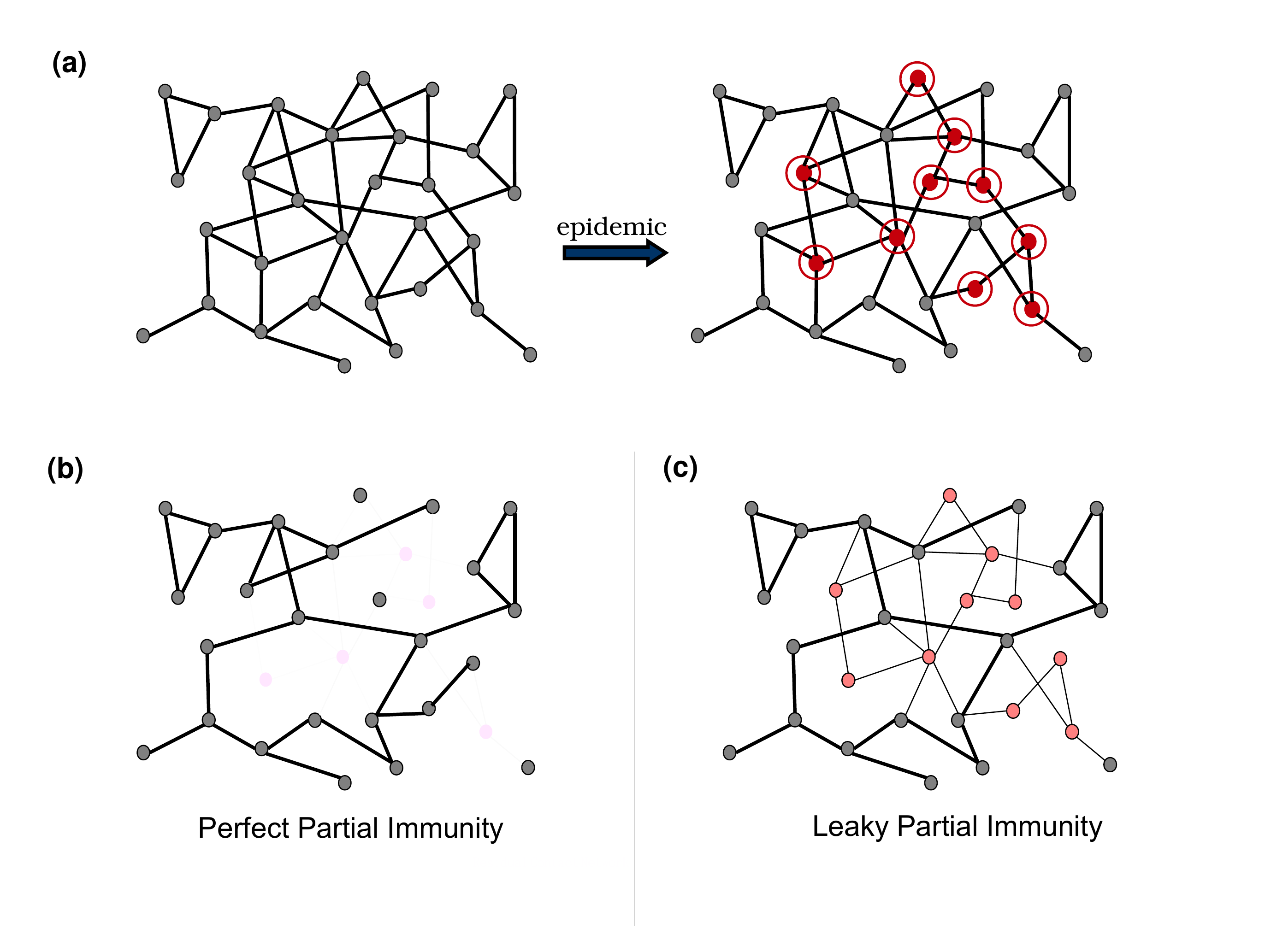}
\par\end{centering}

\caption{Epidemiological contact networks. (a) Prior to an initial epidemic,
all individuals are fully susceptible to disease (gray nodes). Then
some individuals become infected during the epidemic (red nodes).
(b) Perfect partial immunity (at 50\%) means that half of the previously
infected individuals are fully protected against reinfection, while
the other half are fully susceptible again. (c) Leaky partial immunity
(at 50\%) means that all nodes remain in the network, but the edges
leading to and/or from previously infected individuals are half as
likely to transmit disease (illustrated here with the lighter edges.)}

\label{fig:network}
\end{figure}

We present two mathematical approaches to modeling partial immunity.
First, we model perfect partial immunity by completely removing a
fraction of the individuals (their nodes and edges) who are infected
during an epidemic (Figure \ref{fig:network}(b)) from the network.Using
the bond percolation model, we then derive epidemiological quantities
for a subsequent outbreak in the immunized population. Second, we
model leaky partial immunity using a new two-type percolation model.
The underlying contact network topology remains intact, but nodes
are classified either as partially immune or susceptible (Figure \ref{fig:network}(c)).
In both models, we assume that both infectivity and susceptibility
are reduced due to immunity, but the leaky partial immunity model
can be easily adapted to model other effects of immunity. 

Below, we use both models to consider dynamics in three network types:
(a) Poisson, with degree distribution $p_k = e^{-\lambda}\lambda^k/k!$;
(b) exponential, with degree distribution $p_k = (1-e^{\kappa})e^{-\kappa (k-1)}$;
and (c) scale-free, with degree distribution $p_k = k^{-\gamma}/\zeta(\gamma)$,
each with a mean degree of 10. All model predictions are verified
using stochastic simulations which assume a simple percolation process
with parameters to match the model.

\subsection{Perfect Partial Immunity}

Perfect partial immunity, sometimes known as {}``all-or-nothing''
partial immunity or polarized immunity, implies that for a partial
immunity level $\left(1-\alpha\right)$, a fraction $\left(1-\alpha\right)$
of the infected population are fully immune to reinfection (and thus
transmitting to others) and the remaining proportion $\alpha$ are
fully vulnerable to reinfection (and transmission to others thereafter.)
In terms of a contact network, this means that a fraction of the previously
infected nodes are now completely removed (along with its edges) from
the contact network and are no longer a part of the transmission process.
The residual network, introduced in \cite{ferrari,bansal_residual}
models this phenomenon. Previously, we characterized the residual
network as the network made up of uninfected individuals and the edges
connecting them, as we assumed that all infected individuals had gained
full immunity to infection and thus could be fully removed (along
with their edges) from the transmission chain of future epidemics.
Now, we extend the description of the residual network to include
not only uninfected nodes, but also nodes that were previously infected
but have already lost immunity. We apply bond percolation methods
to this extended residual network to model the spread of a subsequent
outbreak in a population that has already suffered an initial outbreak.

The simple Susceptible-Infectious-Recovered (SIR) bond percolation
model allows us to derive fundamental epidemiological quantities based
on the average transmissibility $T$ of the pathogen (that is, the
average probability that an infected node will transmit to a susceptible
contact sometime during its infectious period) and the degree distribution
of the host contact network, denoted $\{p_{k}\}$ where $p_{k}$ is
the fraction of nodes with degree $k$ \cite{mejn}. This assumes
that the probabilities of transmission from infected nodes to susceptible
nodes are \textit{iid} random variables. We can then calculate the
epidemic threshold for a given network ($T_{c})$, above which a large
scale epidemic is possible; this is closely related to the traditional
epidemiological quantity, $R_{0}$. We can also find the probability
and expected size of an epidemic above that threshold as well as the
probability that an individual at the end of a randomly chosen edge
(contact) does not become infected during an epidemic ($u$) \cite{mejn}.
We will apply this method to calculate epidemic quantities for two
consecutive seasons, and use subscripts $1$ and $2$ to denote initial
and subsequent outbreak, respectively. Specifically, $T_{1}$ and
$T_{2}$ denote the average transmissibilities of the pathogen in
each season, respectively, and allow for evolution of infectiousness
from one season to the next; $p_{1}(k)$ and $p_{2}(k)$ denote the
fraction of nodes with $k$ susceptible contacts prior to the first
and second seasons, respectively; and $u_{1}$ and $u_{2}$ denote
the fraction of contacts that remain uninfected following the each
outbreak.

The probability that an individual of degree $k$ will remain uninfected
after the first epidemic can be calculated as $\left(1-T_{1}+T_{1}u_{1}\right)^{k}$
\cite{meyers_sars}. We denote this probability $\eta_{1}(k)$. We
next derive the degree distribution of the epidemiologically active
portion of the network following the initial outbreak. This includes
both nodes that were not infected and nodes that were infected and
subsequently lost immunity, as well as all edges connecting them.
The fraction of \textit{active} nodes with $k$ \textit{active} edges
just prior to the second outbreak is given by

\begin{equation}
p_{2}\left(k\right)=\frac{p_{2}^{uninfected}\left(k\right)+\alpha p_{2}^{infected}\left(k\right)}{\sum\limits _{j}p_{1}(j)\eta_{1}(j)+\alpha\sum\limits _{j}p_{1}(j)\left(1-\eta_{1}(j)\right)}\label{eq:1}\end{equation}
where $p_{2}^{uninfected}\left(k\right)$ and $p_{2}^{infected}\left(k\right)$
are the fractions of susceptible nodes with $k$ susceptible neighbors
among previously uninfected and infected nodes, respectively and $\alpha$
is the proportion of infected individuals who have lost immunity prior
to the second outbreak. The denominator of Equation \ref{eq:1} gives
the proportion of the network that is susceptible prior to the second
outbreak, where the first term considers previously uninfected nodes,
and the second term gives the proportion $\alpha$ of previously infected
nodes.

The probability that a node in the residual network has $k$ remaining
edges (i.e. edges that connect them to other susceptible nodes), given
that it had $\kappa$ edges in the initial network is the following:

\[
p_{2}(k\vert k_{init}=\kappa)=\left(\begin{array}{c}
\kappa\\
k\end{array}\right)\left(u_{1}+(1-u_{1})\alpha\right)^{\kappa}\left((1-u_{1})(1-\alpha)\right)^{\kappa-k}\]
For every node in the residual network, remaining edges include (a)
those that lead to nodes that were uninfected in the previous epidemic
(which occurs with probability $u_{1}$\cite{bansal_residual}) and
(b) those that lead to nodes that were infected but have lost immunity
(which occurs with probability $\left(1-u_{1}\right)\alpha$). Then
the degree distribution prior to season two can thus be rewritten
as,

\[
p_{2}\left(k\right)=\frac{\sum\limits _{\kappa\ge k}p_{1}(\kappa)\eta_{1}(\kappa)p_{2}\left(k|k_{init}=\kappa\right)+\alpha\sum\limits _{\kappa\ge k}p_{1}(\kappa)\left(1-\eta_{1}(\kappa)\right)p_{2}\left(k|k_{init}=\kappa\right)}{\sum\limits _{j}p_{1}(j)\eta_{1}(j)+\alpha\sum\limits _{j}p_{1}(j)\left(1-\eta_{1}(j)\right)}\]
We provide the full derivation of this equation in the Supplementary
Information.

The residual degree distribution $\{p_{2}(k)\}$ reflects the epidemiologically
active portion of the population following the initial epidemic. Although
the residual network differs from the original contact network in
degree distribution, component structure and other topological characteristics,
it is still reasonable to model it as a semi-random graph (as shown
in \cite{bansal_residual}) and thus apply bond percolation methods
\cite{mejn}. Additionally, we show in Supplementary Information that
both immunity models also perform well on non-random realistic or
empirical networks. We next derive epidemiological quantities that
predict the fate of a subsequent outbreak through the residual network.

The probability generating function (PGF) for the second season degree
distribution in terms of the PGF for the initial degree distribution,
$\Gamma_{1}\left(x\right)$ is given by \[
\Gamma_{2}(x)=\frac{\Gamma_{1}\left(r\left(x\left(1-s\right)+s\right)\right)+\alpha\Gamma_{1}\left(\left(1-r\right)\left(x\left(1-s\right)+s\right)\right)}{\Gamma_{1}\left(r\right)+\alpha\left(1-\Gamma_{1}\left(r\right)\right)}.\]
where $r=\left(1-T_{1}+T_{1}u_{1}\right)$ is the probability that
disease was not transmitted along a uniform random edge in the first
epidemic; and $s=\left(1-u_{1}\right)\left(1-\alpha\right)$ is the
probability that a node at the end of a uniform random edge was infected
gained full immunity. 

This allows us to derive the epidemic threshold for the subsequent
outbreak, that is, the critical value of transmissibility above which
a second epidemic is possible, given that some previously infected
individuals have perfect immunity. It is a function of the original
network topology (via the PGF $\Gamma_{1}\left(x\right)$) and the
loss of immunity, $\alpha$, and is given by \[
\begin{array}{c}
\left(T{}_{2_{c}}\right)_{perfect}=\frac{\Gamma_{2}\prime\left(1\right)}{\Gamma_{2}\prime\prime\left(1\right)}=\frac{\Gamma_{1}^{\prime}\left(r\right)r\left(1-s\right)+\alpha\Gamma_{1}^{\prime}\left(1-r\right)\left(1-r\right)\left(1-s\right)}{\Gamma_{1}^{\prime\prime}\left(r\right)r^{2}\left(1-s\right)^{2}+\alpha\Gamma_{1}^{\prime\prime}\left(1-r\right)\left(1-r\right)^{2}\left(1-s\right)^{2}}\end{array}\]
where $\Gamma_{1}^{\prime}\left(r\right),\Gamma_{1}^{\prime}\left(1-r\right)$
are the average degrees among previously uninfected nodes and infected
nodes, respectively. If the second strain is above this epidemic threshold,
then the following equation gives the expected fraction of the residual
population infected during the resulting epidemic

\[
S_{2}=1-\Gamma_{2}\left(u_{2}\right)\]
where $u_{2}$ is the probability that a random edge in the residual
network leads to a node which was uninfected in the second outbreak.
(See Supplementary Information.) Thus the overall fraction expected
to become infected during a second epidemic, assuming perfect partial
immunity at a level $\left(1-\alpha\right)$ is given by \[
\left(S_{2}\right)_{perfect}=S_{2}\left(\sum_{k}p_{1}(k)\eta_{1}(k)+\alpha\left(1-\sum_{k}p_{1}(k)\eta_{1}(k)\right)\right)\]
where $\sum p_{k}\eta_{k}$ represents the size of the population
which was uninfected in the previous outbreak and $\alpha\left(1-\sum p_{k}\eta_{k}\right)$
is the proportion of the population that was infected in the previous
outbreak but has lost immunity.

\subsection{Leaky Partial Immunity}

To model leaky partial immunity, we reduce the probabilities of reinfection
and transmission for nodes infected in the first epidemic. Rather
than deleting nodes and attached edges entirely (as above), we introduce
a two-type percolation approach in which the parameters of disease
transmission depend on the epidemiological history of both nodes involved
in any contact.

\subsubsection{Two-type Percolation}

The standard bond percolation model of \cite{mejn} assumes that,
all nodes of a given degree $k$ are homogeneous with respect to disease
susceptibility and all edges are homogeneous (probabilities of transmission
along edges are i.i.d. random variables with mean $T$). We extend
the basic model to allow for two types of nodes, we call them $A$
and $B$; and four types of edges, $AA,\: AB,\: BA,\: BB$, connecting
all combinations of nodes. (A similar model was recently introduced
in \cite{babak_multitype}.) We use $p_{ij}$ to denote the joint
probability that a uniform random type $A$ node has $i$ edges leading
to other type $A$ nodes and $j$ edges leading to type $B$ nodes
(where $i$ the $A$-degree of the node and $j$ the $B$-degree of
the node). Similarly, $q_{ij}$ denotes the joint probability of a
type $B$ node having an $A$-degree of $i$ and a $B$-degree of
$j$. The multivariate probability generating functions (PGFs) for
these probability distributions are given by \[
f_{A}(x,y)=\sum p_{ij}x^{i}y^{j}\]
\[
f_{B}(x,y)=\sum q_{ij}x^{i}y^{j}\]

\begin{figure}[h]
\begin{centering}
\includegraphics[width=12cm]{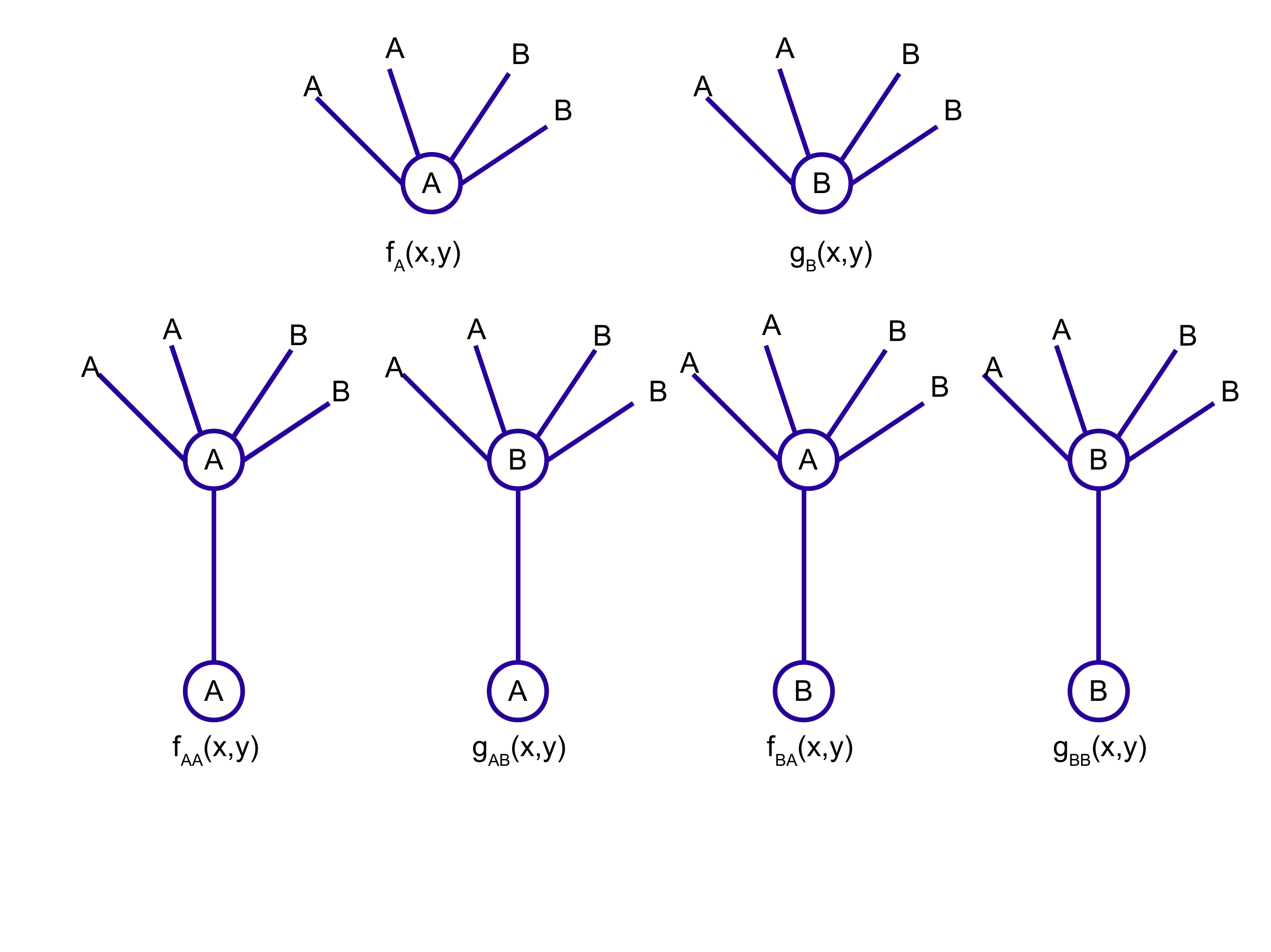}
\par\end{centering}

\caption{The probability generating functions give the numbers of A and B contacts
for each type of vertex (top). The four excess degree distributions
give the numbers of each type of contact for a vertex chosen by following
a uniform random edge (bottom).}

\label{fig:diag}
\end{figure}

While $f_{A}$ and $f_{B}$ describe the distribution of degrees of
randomly chosen $A$ and $B$ nodes, the degree of a node reached
by following a randomly chosen edge is measured by the its excess
degree \cite{mejn}. The PGFs for the $A$-excess degree and the $B$-excess
degree of $A$ and $B$ nodes are given by

\begin{eqnarray*}
f_{AA}(x,y)=\frac{\sum ip_{ij}x^{i-1}y^{j}}{\sum iq_{ij}} &  & f_{BA}(x,y)=\frac{\sum jp_{ij}x^{i}y^{j-1}}{\sum jq_{ij}}\\
f_{AB}(x,y)=\frac{\sum iq_{ij}x^{i-1}y^{j}}{\sum iq_{ij}} &  & f_{BB}(x,y)=\frac{\sum jq_{ij}x^{i}y^{j-1}}{\sum jq_{ij}}\end{eqnarray*}
as illustrated in Figure \ref{fig:diag}.

Having formalized the structure of the contact network in PGFs, we
can now derive the distributions for the number of infected edges,
which are edges over which disease has been successfully transmitted.
We assume that for each edge type ($XY$), transmission probabilities
are i.i.d. random variables with averages denoted $T_{XY}$, and that
these values can vary among the four edge types. Then the PGFs for
the number of occupied edges emanating from a node of type $A$ and
$B$ are, respectivel\textcolor{black}{y:} \[
f_{A}(x,y;T_{AA},T_{AB})=f_{A}((1+(x-1)T_{AA}),(1+(y-1)T_{AB})\]
 \[
f_{B}(x,y;T_{BA},T_{BB})=f_{B}((1+(x-1)T_{BA}),(1+(y-1)T_{BB})\]

Each of these generating functions was derived following the arguments
outlined in \cite{mejn} for the simple bond percolation SIR model.
We can similarly derive the PGFs for the number of infected excess
edges emanating from a node of type $A$ ($B$), at which we arrived
by following a uniform random edge from a node of type $A$ ($B$):
\[
f_{AA}(x,y;T_{AA},T_{AB})=f_{AA}((1+(x-1)T_{AA}),(1+(y-1)T_{AB}))\]
 \[
f_{BA}(x,y;T_{AA},T_{AB})=f_{BA}((1+(x-1)T_{AA}),(1+(y-1)T_{AB}))\]
 \[
f_{AB}(x,y;T_{BA},T_{BB})=f_{AB}((1+(x-1)T_{BA}),(1+(y-1)T_{BB}))\]
 \[
f_{BB}(x,y;T_{BA},T_{BB})=f_{BB}((1+(x-1)T_{BA}),(1+(y-1)T_{BB}))\]

The PGFs for outbreak sizes starting from a node of type $A$ or $B$,
respectively, are then given by \[
F_{A}(x,y;T_{AA},T_{AB})=xf_{A}(F_{AA}(x,y;\left\{ T\right\} ),F_{AB}(x,y;\left\{ T\right\} );T_{AA},T_{AB})\]
 \[
F_{B}(x,y;T_{BA},T_{BB})=yf_{B}(F_{BA}(x,y;\left\{ T\right\} ),F_{BB}(x,y;\left\{ T\right\} );T_{BA},T_{BB})\]

where, $F_{AA}$ and $F_{BA}$ are the PGFs for the outbreak size
distribution starting from an (infected) node of type $A$ which has
been reached by following an edge from another (infected) node of
type $A$ or $B$, respectively. Similarly, $F_{AB}$ and $F_{BB}$
are the PGFs for the outbreak size distribution starting from an (infected)
node of type $B$ which has been reached by following an edge from
another (infected) node of type $A$ or $B$, respectively. These
PGFs are as follows

\[
F_{AA}(x,y;\left\{ T\right\} )=xf_{AA}(F_{AA}(x,y;\left\{ T\right\} ),F_{AB}(x,y;\left\{ T\right\} );T_{AA},T_{AB})\]
 \[
F_{BA}(x,y;\left\{ T\right\} )=xf_{BA}(F_{BA}(x,y;\left\{ T\right\} ),F_{BB}(x,y;\left\{ T\right\} );T_{AA},T_{AB})\]
 \[
F_{AB}(x,y;\left\{ T\right\} )=yf_{AB}(F_{BA}(x,y;\left\{ T\right\} ),F_{BB}(x,y;\left\{ T\right\} );T_{BA},T_{BB})\]
 \[
F_{BB}(x,y;\left\{ T\right\} )=yf_{BB}(F_{BA}(x,y;\left\{ T\right\} ),F_{BB}(x,y;\left\{ T\right\} );T_{BA},T_{BB})\]

Again following the method of \cite{mejn}, we can derive the expected
size of a small outbreak and the epidemic threshold (given in the
Supplementary Information). The expected numbers of $A$ and $B$
nodes infected in a small outbreak are found by taking partial derivatives
of the PGF for the outbreak size distribution: \[
\left\langle s\right\rangle _{A}=\frac{\partial F_{A}}{\partial x}|_{x=1,y=1}+\frac{\partial F_{B}}{\partial x}|_{x=1,y=1}\]

\[
\left\langle s\right\rangle _{B}=\frac{\partial F_{A}}{\partial y}|_{x=1,y=1}+\frac{\partial F_{B}}{\partial y}|_{x=1,y=1}\]

Finally, we can find the size of a large-scale epidemic among $A$
nodes and among $B$ nodes as: \begin{equation}
S_{A}\left(T_{AA},T_{AB}\right)=1-F_{A}(1,1;T_{AA},T_{AB})=1-\sum p_{ij}(1+(a-1)T_{AA})^{i}(1+(c-1)T_{AB})^{j}\label{eq:4}\end{equation}
 \begin{equation}
S_{B}\left(T_{BA},T_{BB}\right)=1-F_{B}B(1,1;T_{BA},T_{BB})=1-\sum q_{ij}(1+(b-1)T_{BA})^{i}(1+(d-1)T_{BB})^{j}\label{eq:5}\end{equation}

where, $a=F_{AA}(1,1;\left\{ T\right\} ),b=F_{BA}(1,1;\left\{ T\right\} ),c=F_{AB}(1,1;\left\{ T\right\} ),d=F_{BB}(1,1;\left\{ T\right\} )$.
The probability of a large-scale epidemic can be derived similarly.
The numerical values for the size and probability of an outbreak will
be equal if $T_{AB}=T_{BA}$. Further details are provided in the
Supplementary Information.

This two-type percolation model provides a general framework for modeling
pathogens with variable transmissibility and host populations with
immunological heterogeneity.

\subsubsection{Modeling Leaky Immunity with Two-Type Percolation }

We now apply the two-type percolation method to model leaky partial
immunity. In this model, type $A$ nodes represent individuals who
were not infected in the initial epidemic and thus have no prior immunity,
and type $B$ nodes represent those who were infected and maintain
partial immunity (at a level $1-\alpha$). (Note that $\alpha$ gives
the fraction of immunity lost in both models.) Here, we assume that
prior immunity causes equal sized reductions in both infectivity and
susceptibility ($\alpha$); but the approach can be extended easily
to include more complex models of immunity. Specifically, during the
subsequent epidemic, type A individuals (previously uninfected) have
a susceptibility of one and an infectivity of $T_{2},$while type
B individuals (previously infected) have a susceptibility of $\alpha$
and an infectivity of $T_{2}\alpha$. Correspondingly, $T_{AA}=T_{2}$,
$T_{AB}=T_{2}\alpha$, $T_{BA}=T_{2}\alpha$, and $T_{BB}=T_{2}\alpha^{2}$
. 

The joint degree distributions for type A and type B nodes depend
on the course of the initial epidemic, and are given by

\[
p_{ij}=p_{1}(i+j)\eta_{1}(i+j)\left(\begin{array}{c}
i+j\\
i\end{array}\right)u_{1}^{i}\left(1-u_{1}\right)^{j}\]

\[
q_{ij}=p_{1}(i+j)\left(1-\eta_{1}(i+j)\right)\left(\begin{array}{c}
i+j\\
i\end{array}\right)u_{1}^{i}\left(1-u_{1}\right)^{j}\]

respectively, where $i$ is the A-degree and $j$ is the B-degree.
Further explanation can be found in Supplementary Information. 

Using the quantities derived above, we can model epidemics that leave
varying levels of individual-level partial immunity. Using equations
\ref{eq:4} and \ref{eq:5}, for example, we can solve for the size
of the epidemic in a second epidemic with (individual-level) leaky
partial immunity, $\left(1-\alpha\right)$:

\[
\left(S_{2}\right)_{leaky}=\left(\sum_{k}p_{1}(k)\eta_{1}(k)\right)S_{A}\left(T_{2},T_{2}\alpha\right)+\left(1-\sum_{k}p_{1}(k)\eta_{1}(k)\right)S_{B}\left(T_{2}\alpha,T_{2}\alpha^{2}\right).\]

\section{Results}

\subsection{Impact of One Epidemic on the Next}

We have introduced two distinct mathematical approaches for modeling
the epidemiological consequences of naturally-acquired immunity. The
residual network model probabilistically removes nodes and edges corresponding
to the fraction ($\alpha$) of infected nodes expected to lose immunity
entirely. The two-type percolation model tracks the epidemiological
history of all individuals and reduces the infectivity and susceptibility
of all previously infected nodes by a fraction ($\alpha$). By adjusting
$\alpha$, both models can explore the entire range of immunity from
none to complete. At $\alpha=1$ , these model the absence or complete
loss of immunity and thus would apply when the second season strain
is entirely antigenically distinct from the prior strain. At $\alpha=0$,
these model full or no loss of prior immunity and might apply when
a secondary epidemic is caused by the same or very similar pathogen
as caused the first epidemic. Values of $\alpha$ between 0 and 1
represent partial immunity to the second pathogen, with the level
of protection increasing as $\alpha$ approaches 0.

\begin{figure}[h]
\begin{centering}
\includegraphics[width=14cm]{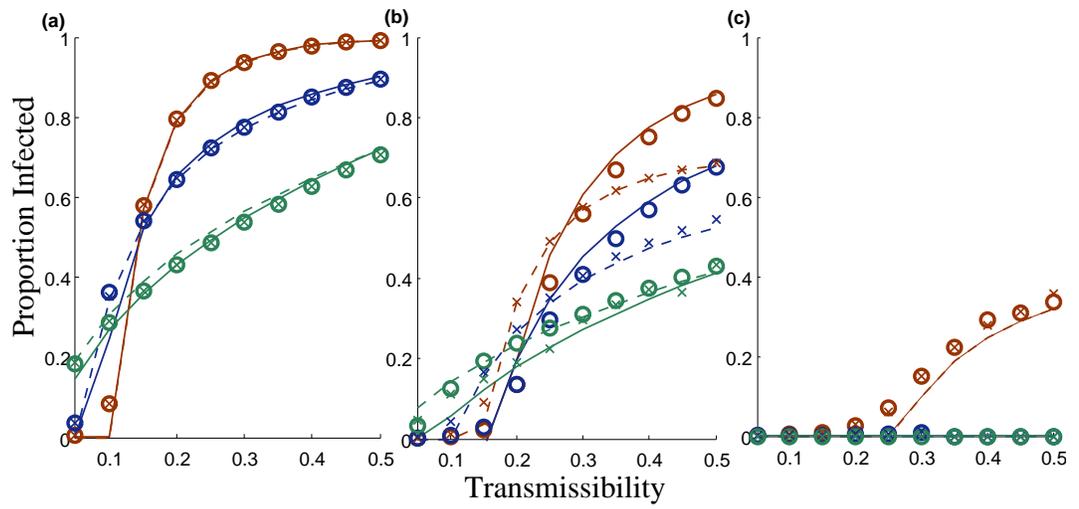}
\par\end{centering}

\caption{Expected size of a second epidemic as infectiousness increases. We
compare the predictions of our mathematical models for perfect (dashed
line) and leaky (solid line) immunity to corresponding numerical simulations
(crosses and circles indicate perfect and leaky immunity, respectively).
Calculations are for three types of networks: Poisson (red), exponential
(blue), and power law (green) with mean degree 10, for three levels
of immunity: (a) no immunity ($\alpha=1$), (b) partial immunity ($\alpha=0.5$),
and (c) full immunity ($\alpha=0$), and for a range of second strain
transmissibility values ($T_{2}$) along each x-axis (assuming $T_{1}=0.15$
in all cases). }

\label{fig:compare_imm}
\end{figure}

In Figure \ref{fig:compare_imm}, we compare the predicted sizes of
a second epidemic for both the perfect and leaky models against simulations
for a Poisson, exponential and scale-free random network (of the same
mean degree) and under the conditions of no prior immunity ($\alpha=1$),
partial immunity ($\alpha=0.5$), and full immunity ($\alpha=0$)
for values of transmissibility between $0$ and $0.5$ . T\textcolor{black}{here
is a strong congruence between our analytical calculations and their
corresponding simulations.} Assuming no immunity (Figure \ref{fig:compare_imm}(a)),
the two models simplify to the standard bond percolation model on
the original network, and thus make identical predictions. Assuming
full immunity (Figure \ref{fig:compare_imm}(c)), the perfect immunity
model removes all previously infected nodes (and the corresponding
edges) before the second outbreak; and the leaky immunity model sets
transmissibility along all edges leading from and to previously infected
nodes to zero, thus de-activating those nodes entirely. Consequently,
the models also converge at this extreme. The two models are, however,
fundamentally different for any level of intermediate partial immunity
between $\left({0<\alpha<1}\right)$ as they assume different models
of immunological protection (Figure \ref{fig:compare_imm}(b)). At
$\alpha=0.5$, leaky immunity confers greater herd immunity than perfect
immunity at low values of transmissibility, while the reverse is true
for more infectious pathogens. The makeup of the previously infected
population is identical in both models and biased towards high degree
individuals. When the pathogen is only mildly contagious, leaky immunity
goes a long way towards protecting all previously infected hosts whereas
perfect immunity protects only a fraction of these hosts; when it
is more highly contagious, however, leaky immunity is insufficient
to protect hosts with large numbers of contacts whereas perfect immunity
is not diminished. We find further that network heterogeneity acts
consistently across different levels of immunity. The Poisson network
has the most homogeneous degree distribution followed by the exponential
network and finally the scale-free network with considerable heterogeneity.
Holding mean degree constant, variance in degree increases the vulnerability
of the population (allowing epidemics to occur at lower rates of transmissibility),
yet generally reduces the ultimate size of epidemics when they occur.
At high levels of immunity, the susceptible network at the start of
the second season becomes more sparse and homogeneous. Thus the impact
of network variance on the second epidemic diminishes as immunity
increases, that is, as $\alpha\rightarrow0$. (We elaborate further
on these results in the Supplementary Information.) 

\begin{figure}[h]
\begin{centering}
\includegraphics[width=14cm]{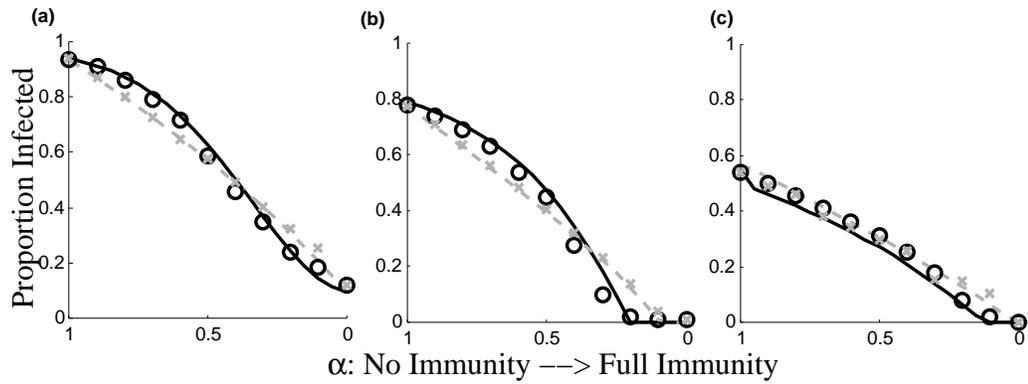}
\par\end{centering}

\caption{Expected size of a second epidemic as immunity increases. We compare
predictions of the perfect immunity model (gray dashed lines), leaky
immunity model (black solid lines) and simulations for each model
(gray cross and black circle markers, respectively). Calculations
and simulations are for networks with (a) Poisson, (b) exponential,
and (c) scale-free degree distributions with mean degree 10, at transmissibilities
$T_{1}=0.15$ and $T_{2}=0.3$.}

\label{res-fig7} 
\end{figure}

We explore intermediate levels of immunity further in Figure \ref{res-fig7},
and again find reasonable agreement between our analytic predictions
and simulations. As expected, increasing levels of immunity (from
left to right) decrease the epidemic potential of a second outbreak.\textcolor{red}{
}At these intermediate values of transmissibility ($T_{1}=0.15$ and
$T_{2}=0.3$), leaky immunity tends to confer lower herd immunity
than perfect immunity, except at extremely high levels of immunity.
The level of immunity at which the predicted epidemic sizes for two
immunity models cross represents the point at which leaky partial
immunity for all prior cases effectively protects more individuals
than the complete removal of a fraction of those cases. This transition
point occurs at a higher level of immunity in the exponential network
than the Poisson network, and never occurs in the scale-free network,
perhaps because the immunized individuals in the more heterogeneous
networks tend to have anomalously high numbers of contacts thus limiting
the efficacy of partial protection.

\subsection{Pathogen Re-invasion and Immune Escape}

When a pathogen enters a population that has experienced a prior outbreak,
its success will depend on the extent and pattern of naturally-acquired
immunity in the host population. The new pathogen may not be able
to invade unless it is significantly different from the original strain.
If it is antigenically distinct from the prior strain, then prior
immunity may be irrelevant; and if it is more transmissible than the
original strain, then it may have the potential to reach previously
unexposed individuals.

\begin{figure} [h]
\centering{}\includegraphics[width=14cm]{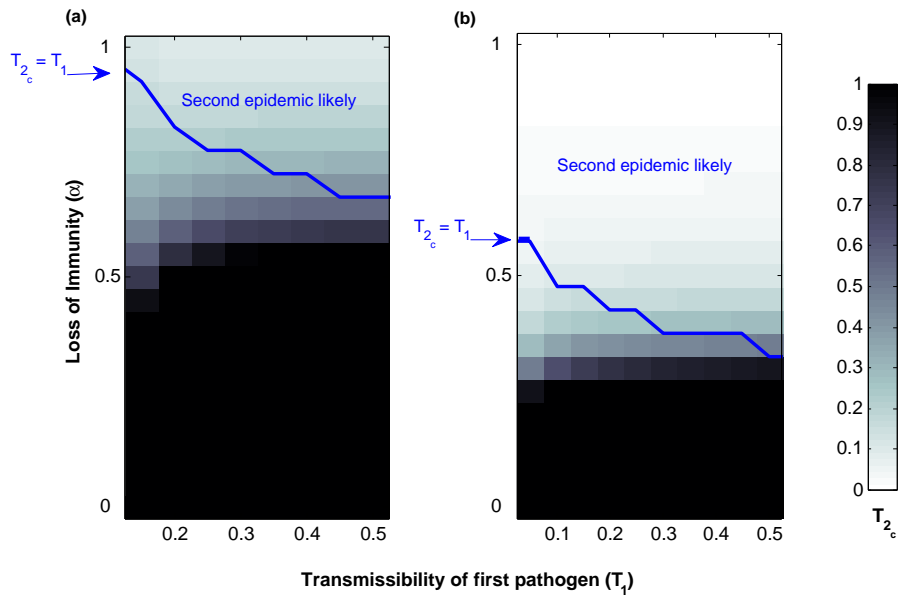}\caption{Epidemic threshold ($T_{2_{c}})$ in the second season. The colors
indicate the level of transmissibility required for the second strain
to invade the population (cause an epidemic), assuming leaky partial
immunity for (a) a Poisson-distributed network and (b) a scale-free
network, each with mean degree of 10. The x-axis gives the first season
transmissibility ($T_{1}$) and y-axis gives the loss of immunity
($\alpha$). The blue line denotes $T_{2_{c}}=T_{1}$,: above the
line $T_{2_{c}}<T_{1},$ and invasion by the original pathogen is
possible.\label{fig:reinvasion}}

\end{figure}

Figure \ref{fig:reinvasion} indicates the minimum transmissibility
required for the new strain to cause an epidemic (that is, its critical
transmissibility $T_{2_{c}}$), as a function of the transmissibility
of the original strain ($T_{1}$) and the level of leaky immunity
($\alpha$). The leakier the immunity (high $\alpha$) and the lower
the infectiousness of the original strain (low $T_{1}$), the more
vulnerable the population to a second epidemic (light coloration in
Figure \ref{fig:reinvasion}). Generally the homogeneous Poisson network
is less vulnerable to re-invasion than the heterogeneous scale-free
network. The blue curves in figure \ref{fig:reinvasion} show combinations
of $T_{1}$ and $\alpha$ where the epidemic threshold for the new
strain equals the transmissibility of the original strain ($T_{2_{c}}=T_{1}$)
and have two complementary interpretations. First, if we assume that
the new strain is exactly as transmissible as the original strain
($T=T_{2}=T_{1}$), then the curves indicate the critical level of
cross-immunity ($\alpha_{c}(T)$) below which the strain can never
invade and above which the strain can invade with some probability
that increases with $\alpha$. This threshold indicates the extent
of antigenic evolution (or intrinsic decay in immune response) required
for a second epidemic to occur. The more heterogeneous the contact
patterns (scale-free versus Poisson network), the lower the amount
of immune escape required for a pathogen of the same transmissibility
to re-invade. Second, if we assume a fixed level of immune decay ($\alpha$),
then the curves indicate the critical initial transmissibility ($T_{1}$)
above which the new strain can only invade if it is more contagious
than the original strain ($T_{2}>T_{1}$). Below this point, the network
topology and preexisting immunity create a selective environment that
excludes the original strain and favors more transmissible variants.
The perfect immunity model yields similar results (Supplementary Information.)

Some epidemiologists have speculated that there are 'trade-offs' between
virulence and infectiousness, implying that more infectious pathogens
will necessarily be more virulent \cite{anderson_82,ewald,bull,frank}.
If true, figure \ref{fig:reinvasion} suggests that naturally-acquired
immunity, by opening niches for more infectious variants, may indirectly
lead to the evolution of greater virulence. This is consistent with
a previous study showing that that host populations with high levels
of immunity maintain more virulent pathogens than na�ve host populations
\cite{gandon}.

\section{Discussion \& Conclusion}

In this work, we have considered the impact of pathogen spread on
future outbreaks of the same or similar pathogen and on pathogen invasion
and evolution. We have compared two standard models for immunity,
perfect and leaky, and found that the extent of herd immunity varies
with the pathogen transmissibility and the degree and nature of immunity.
Leaky immunity appears to confer greater herd immunity at moderate
levels of pathogen infectiousness for all levels of partial immunity,
whereas perfect immunity is more effective at higher transmissibilities. 

This analysis also has implications for public health intervention
strategies. Contact-reducing interventions (e.g., patient quarantine
and social distancing) and vaccination often result in complete removal
of a fraction of individuals from the network (akin to perfect immunity),
whereas transmission-reducing interventions (e.g., face-masks and
other hygienic precautions) typically reduce transmissibility along
edges leading to and from a fraction of individuals (akin to leaky
immunity) \cite{pourbohloul}. These results thus suggest that contact
reductions will be more effective than a comparable degree of transmission
reductions at higher levels of pathogen infectiousness.

The evolution of new antigenic characteristics in a pathogen that
escape prior immunity and the evolution of higher transmissibility
both depend on genetic variation. Thus, the more infections there
are in the first season, the greater the opportunity for evolutionary
change \cite{boni}. This poses a trade-off for the pathogen: a large
initial epidemic may generate variation that fuels evolution yet wipes
out the susceptible pool for the subsequent season; while a small
initial outbreak leaves a large fraction of the network susceptible
to future transmission yet may fail to generate sufficient antigenic
or other variation for future adaptation. We have shown that the trade-off
between generating immunity via infections and escaping immunity via
antigenic drift will depend not only on the size of the susceptible
population, but also on its connectivity. Although we have focused
primarily on the role of antigenic drift, these models also apply
to loss of immunity through decay in immunological memory, as occurs
following pertussis and measles infections \cite{mossong_measles,pertussis}.

Much epidemiological work, particularly the analysis of intervention
strategies, ignores the immunological history of the host population.
Thus our effort to incorporate host immune history into a flexible
individual-based network model will potentially advance our understanding
of the epidemiological and evolutionary dynamics of partially-immunizing
infections such as influenza, pertussis, or rotavirus. However, these
provide just an initial step in this direction, as the models consider
only two consecutive seasons and do not yet allow for replenishment
or depletion of susceptibles due to births and deaths.

\section*{Acknowledgments}

This work was supported by the RAPIDD program of the Science \& Technology
Directorate, Department of Homeland Security, and the Fogarty International
Center, National Institutes of Health, and grants from the James F.
McDonnell Foundation and National Science Foundation (DEB-0749097)
to L.A.M.

\bibliographystyle{plain}
\bibliography{ms1_with_figs}

\begin{thebibliography}{10}

\bibitem{babak_multitype}
A.~Allard, P.~Noel, L.~Dube, and B.~Pourbohloul.
\newblock Heterogeneous bond percolation on multitype networks with an
  application to epidemic dynamics.
\newblock {\em Phys Rev E}, 79 (3):036113, 2009.

\bibitem{anderson_82}
R.~M. Anderson and R.~M May.
\newblock Coevolution of hosts and parasites.
\newblock {\em Parasitology}, 85:411--426, 1982.

\bibitem{bansal_residual}
S.~Bansal, M.~Ferrari, and L.A. Meyers.
\newblock The structural evolution of host populations and the feedback on
  pathogens.
\newblock {\em In preparation.}

\bibitem{bansal_interface}
S.~Bansal, B.~Grenfell, and L.A. Meyers.
\newblock When individual behavior matters.
\newblock {\em J. R. Soc. Interface}, 4(16), 2007.

\bibitem{barbour}
A.~Barbour and D.~Mollison.
\newblock {\em Stochastic Processes in Epidemic Theory}, chapter Epidemics and
  random graphs, pages 86--89.
\newblock Springer, 1990.

\bibitem{boni}
M.~F. Boni, J.~R. Gog, V.~Andreasen, and F.~B. Christiansen.
\newblock Influenza drift and epidemic size: the race between generating and
  escaping immunity.
\newblock {\em Theoretical population biology}, 65(2):179--191, March 2004.

\bibitem{boni_niche}
M.F. Boni and M.W. Feldman.
\newblock Evolution of antibiotic resistance by human and bacterial niche
  construction.
\newblock {\em Evolution}, 59 (3), 2005.

\bibitem{boots_sasaki}
M.~Boots and A.~Sasaki.
\newblock Small worlds and the evolution of virulence.
\newblock {\em Proc. R. Soc. B}, 266:1933--1938, 1999.

\bibitem{buckee_straindiversity}
C.O. Buckee, K.~Koelle, M.J. Mustard, and S.~Gupta.
\newblock The effects of host contact network structure on pathogen diversity
  and strain structure.
\newblock {\em PNAS}, 101 (29):10839--44, 2004.

\bibitem{bukh}
J.~Bukh, R.~Thimme, J.~Meunier, K.~Faulk, H.~Spangenberg, K.~Chang,
  W.~Satterfield, F.~Chisari, and R.~Purcell.
\newblock Previously infected chimpanzees are not consistently protected
  against reinfection or persistent infection after reexposure to the identical
  hepatitis c virus strain.
\newblock {\em Journal of Virology}, 82:8183--8195, 2008.

\bibitem{bull}
J.~Bull.
\newblock Virulence.
\newblock {\em Evolution}, 48:1423--35, 1994.

\bibitem{varicella}
S.~Chaves, P.~Gargiullo, J.~Zhang, R.~Civen, and D.~et~al Guris.
\newblock Loss of vaccine-induced immunity to varicella over time.
\newblock {\em NEJM}, 356:1121--1129, 2007.

\bibitem{flu_immunity_2}
M.L. Clements, R.F. Betts, E.L. Tierney, and B.R. Murphy.
\newblock Serum and nasal wash antibodies associated with resistance to
  experimental challenge with influenza a wild-type virus.
\newblock {\em J Clin Microbiol}, 24:157--60, 1986.

\bibitem{ewald}
P.W Ewald.
\newblock Transmission modes and evolution of the parasitism-mutualism
  continuum.
\newblock {\em Ann. NY Acad. Sci.}, 503:295--306, 1987.

\bibitem{hepc_farci}
P.~{Farci}, H.~J. {Alter}, S.~{Govindarajan}, D.~C. {Wong}, R.~{Engle}, R.~R.
  {Lesniewski}, I.~K. {Mushahwar}, S.~M. {Desai}, R.~H. {Miller}, N.~{Ogata},
  and R.~H. {Purcell}.
\newblock {Lack of Protective Immunity Against Reinfection with Hepatitis C
  Virus}.
\newblock {\em Science}, 258:135--140, October 1992.

\bibitem{ferrari}
M.~Ferrari, S.~Bansal, L.A. Meyers, and O.~Bjornstad.
\newblock Network frailty and the geometry of herd immunity.
\newblock {\em Proc. of R. Soc}, 273:2743--2748, 2006.

\bibitem{frank}
S.A. Frank.
\newblock Models of parasite virulence.
\newblock {\em Q. Rev. Biol.}, 71:37--78, 1996.

\bibitem{gandon}
Sylvian Gandon, Margaret~J. Mackinnon, Sean Nee, and Andrew~F. Read.
\newblock Imperfect vaccines and the evolution of pathogen virulence.
\newblock {\em Nature}, 414:751--756, 2001.

\bibitem{grassly}
N.C. Grassly, C.~Fraser, and G.P. Garnett.
\newblock Host immunity and synchronized epidemics of syphilis across the
  united states.
\newblock {\em Nature}, 433:417--421, 2005.

\bibitem{flu_book}
R.E. Hope-Simpson.
\newblock {\em The transmission of epidemic influenza}.
\newblock Plenum Press, New York, 1992.

\bibitem{hoppen}
F.~Hoppenstead and P.~Waltman.
\newblock A problem in the theory of epidemics ii.
\newblock {\em Math Biosci}, 12:133--145, 1971.

\bibitem{immunology}
S.~Kaufmann, A.~Sher, and R.~Ahmed.
\newblock {\em Immunology of Infectious Diseases}.
\newblock ASM Press, 2002.

\bibitem{kermack}
W.O. Kermack and A.G. McKendrick.
\newblock A contribution to the mathematical theory of epidemics.
\newblock {\em Proc. R. Soc. A}, 115:700--721, 1927.

\bibitem{levin_dushoff_plotkin}
S.~Levin, J.~Dushoff, and J.~Plotkin.
\newblock Evolution and persistence of influenza a and other diseases.
\newblock {\em Math Biosci}, 188:17--28, 2004.

\bibitem{meyers_sars}
L.A. Meyers, B.~Pourbohloul, M.E.J. Newman, D.M. Skowronski, and R.C Brunham.
\newblock Network theory and sars: predicting outbreak diversity.
\newblock {\em J. Theo. Biol}, 232:71--81, 2005.

\bibitem{mossong_measles}
J.~Mossong and C.P.clem Muller.
\newblock Modelling measles re-emergence as a result of waning of immunity in
  vaccinated populations.
\newblock {\em Vaccine}, 21:4597--4603, 2003.

\bibitem{mejn}
M.E.J. Newman.
\newblock Spread of epidemic disease on networks.
\newblock {\em Phys. Rev. E.}, 66(016128), 2002.

\bibitem{nunes_pathdiversity}
A.~Nunes, M.M. Telo~da Gama, and M.G.M Gomes.
\newblock Localized contacts between hosts reduce pathogen diversity.
\newblock {\em Journal of Theoretical Biology}, 241:477--87, 2006.

\bibitem{nuno_flu_models}
M.~Nuno, C.~Castillo-Chavez, Z.~Feng, and M.~Martcheva.
\newblock {\em Lecture Notes in Mathematics}, chapter Mathematical Models of
  Influenza: The Role of Cross-Immunity, Quarantine and Age-Structure, pages
  349--364.
\newblock Springer Berlin, 2008.

\bibitem{niche_feldman}
F.J. Odling-Smee, K.N. Laland, and M.W. Feldman.
\newblock Niche construction: The neglected process in evolution.
\newblock {\em Monographs in Population Biology.}, 2003.

\bibitem{pastor_complex}
R.~Pastor-Satorras and Vespignani A.
\newblock Epidemic dynamics and endemic states in complex networks.
\newblock {\em Phys Rev E}, 63(066117), 2001.

\bibitem{pourbohloul}
B.~Pourbohloul, L.A. Meyers, D.M. Skowronski, M.~Krajden, and D.M. Patrick.
\newblock Modeling control strategies of respiratory pathogens.
\newblock {\em Emerg Infect Dis}, 11:1249--1256, 2005.

\bibitem{read_net_evolution}
J.~Read and M.J Keeling.
\newblock Disease evolution on networks: the role of contact structure.
\newblock {\em Proc R Soc B}, 270:699--708, 2003.

\bibitem{recker_flu_immunity}
M.~Recker, O.~Pybus, S.~Nee, and S.~Gupta.
\newblock The generation of influenza outbreaks by a network of host immune
  responses against a limited set of antigenic types.
\newblock {\em PNAS}, 104:7711--7716, 2007.

\bibitem{shirley}
M.D.F. Shirley and S.P Rushton.
\newblock The impacts of network topology on disease spread.
\newblock {\em Ecol. Complex}, 2:287--299, 2005.

\bibitem{flu_immunity}
J.L.0 Shulman.
\newblock Effects of immunity on transmission of influenza: Experimental
  studies.
\newblock {\em Progr. Med. Virol.}, 12:128--16, 1970.

\bibitem{van_baalen}
M.~van Baalen.
\newblock {\em Adaptive Dynamics of Infectious Diseases: in Pursuit of
  Virulence Management}, chapter Contact networks and the evolution of
  virulence.
\newblock Cambridge University Press, Cambridge, 2002.

\bibitem{pertussis}
M.~van Boven, H.E. de~Melker, J.F. Schellekens, and M.~Kretzschmar.
\newblock Waning immunity and sub-clinical infection in an epidemic model:
  implications for pertussis in the netherlands.
\newblock {\em Mathematical Biosciences}, 164:161--182, 2000.

\bibitem{waltman}
P.~Waltman.
\newblock Deterministic threshold models in the theory of epidemics.
\newblock {\em Lecture Notes in Biomathematics}, 1, 1974.

\bibitem{watts}
D.~Watts and S.H Strogatz.
\newblock Collective dynamics of small world networks.
\newblock {\em Nature}, 393(441), 1998.

\bibitem{rotavirus}
L.~White, J.~Buttery, B.~Cooper, D.~Nokes, and G.~Medley.
\newblock Rotavirus within day care centres in oxfordshire, uk:
  characterization of partial immunity.
\newblock {\em J R Soc Interface}, 5:1481--1490, 2008.

\end{thebibliography}

\pagebreak{}


\end{document}